%% file: note_antenna_selection.tex
\documentclass[10pt,journal,a4paper]{IEEEtran}

\usepackage{graphicx}
\usepackage{amsfonts}
\usepackage{algorithmic}
\usepackage{algorithm}
\usepackage{amssymb}
\usepackage{bbm}
\usepackage{color}
\usepackage{cite}
\usepackage{amsfonts}
\usepackage{subfigure}
\usepackage{relsize}
\usepackage[cmex10]{amsmath}
\usepackage{graphicx}
\usepackage{array}
\usepackage{amssymb}
\usepackage{bbm}
\usepackage{mdwmath}
\usepackage{mdwtab}
\usepackage{fixltx2e}
\usepackage{url}
\usepackage[center,scriptsize]{caption}
\newtheorem{theorem}{Theorem}[section]
\newtheorem{property}{Property}[section]

\usepackage{amsmath,amssymb,mathrsfs}
\usepackage{epsfig,epstopdf,epsf,subfigure,graphicx,graphics,subfigure}
\usepackage{url}
\usepackage{color}

\newcommand{\mb}{\mathbf}

\allowdisplaybreaks
\IEEEoverridecommandlockouts

\title{A Note on Antenna Selection in Gaussian MIMO Channels: Capacity Guarantees and Bounds}
\author{
    \IEEEauthorblockN{Yahya H. Ezzeldin\IEEEauthorrefmark{1}, Ayan Sengupta\IEEEauthorrefmark{1}, Christina Fragouli\IEEEauthorrefmark{1}}
    \IEEEauthorblockA{\\\IEEEauthorrefmark{1}Department of Electrical Engineering\\University of California Los Angeles, USA
    \\\{yahya.ezzeldin, ayansg, christina.fragouli\}@ucla.edu}
}

\begin{document}
\maketitle
\begin{abstract}
    We consider the problem of selecting $k_t \times k_r$ antennas from a Gaussian MIMO channel with $n_t \times n_r$ antennas, where $k_t \leq n_t$ and $k_r \leq n_r$.
    We prove the following two results with regards to the same, that hold \emph{universally}, i.e., these do not depend on the channel coefficients:
    (i) The capacity of the best $k_t \times k_r$ antennas is always lower bounded by a fraction $\frac{k_t\cdot k_r}{n_t \cdot n_r}$ of the full capacity (with $n_t \times n_r$ antennas).
    This bound is tight as the channel coefficients diminish in magnitude.
    (ii) The best $k_t \times k_r$ antennas always achieve a fraction greater than $\frac{\min\{k_t,k_r\}}{\min\{n_t,n_r\}}$ of the full capacity \emph{within an additive constant} that is independent of the channel coefficients. This bound is tight (up to the additive constant) for parallel channels.
The key mathematical idea that allows us to derive these universal bounds is to directly relate the determinants of principal sub-matrices of a Hermitian matrix to the determinant of the entire matrix.
\end{abstract}


\input{mainresults.tex}

\input{proof.tex}

\input{appendices.tex}

\bibliographystyle{IEEEtran}
\bibliography{references}

\end{document}

%% file: mainresults.tex
\section{Problem Formulation and Main Results}
\label{mainresults}
We consider the Gaussian $n_t \times n_r$ MIMO channel with independent and identically distributed (i.i.d) inputs.
Let $\mathbf{x} \in \mathbb{C}^{n_t \times 1}$ denote the transmitted signal from the $n_t$ transmitter antennas and $\mathbf{y} \in \mathbb{C}^{n_r \times 1}$ denote the signal received by the $n_r$ receiver antennas. Then the signal flow through this MIMO channel is given by
\begin{align*}
    \mathbf{y} = \mathbf{H}\mathbf{x} + \mathbf{z}
\end{align*}
where $\mathbf{z} \in \mathbb{C}^{n_r \times 1}$ is a random circularly symmetric complex Gaussian vector with zero mean and identity covariance matrix; $\mathbf{H}$ represents the $n_r \times n_t$ MIMO channel matrix.
The capacity of this MIMO channel, with individual (average) power constraints $P$ at the transmitters is \cite{tsefundamentals}
\[
    C = \log\det(I + P\mathbf{H}\mathbf{H}^\dagger).
\]

Our goal is to evaluate universal guarantees (on capacity) that exist, if one selects the best $k_t \times k_r$ subchannel from the $n_t \times n_r$ channel. The capacity of the best $k_t \times k_r$ subchannel is given by:
\[
    C_{k_t,k_r}^{\max} = \max_{\Lambda}\ \ \log\det(I + P\mathbf{H}_\Lambda {\mathbf{H}_\Lambda}^\dagger),
\]
where $\mathbf{H}_\Lambda$ denotes a $k_r \times k_t$ principal submatrix of $\mathbf{H}$ representing the channel coefficients between the chosen $k_t$ transmitters and $k_r$ receivers.
\\
The following two theorems summarize our main results.
\begin{theorem}
    \label{zero_gap_thm}
    Consider an $n_t \times n_r$ Gaussian MIMO channel with i.i.d inputs, individual power constraint $P$ at the transmit antennas, and capacity $C$. Then there always exists a $k_t \times k_r$ subchannel with capacity $C_{k_t,k_r}^\star$ such that:
    \begin{equation}
        C_{k_t,k_r}^{\max} \geq C_{k_t,k_r}^\star \geq \dfrac{k_t\cdot k_r}{n_t\cdot n_r}\ C.
    \label{zero_gap}
    \end{equation}
    
    Moreover, there exist MIMO channel configurations for which $C_{k_t,k_r}^{\max} = \frac{k_t\cdot k_r}{n_t\cdot n_r}\ C$.
\end{theorem}

\begin{theorem}
    \label{const_gap_thm}
    For every $n_t \times n_r$ Gaussian MIMO channel with i.i.d inputs, individual power constraint $P$ at the transmit antennas, and capacity $C$, there exists a $k_t \times k_r$ subchannel with capacity $C_{k_t,k_r}^\star$ such that:
    \begin{equation}
        C_{k_t,k_r}^{\max} \geq C_{k_t,k_r}^\star \geq \dfrac{\min(k_t,k_r)}{\min(n_t,n_r)}\ C - G.
    \end{equation}
    where $G = \log\left({n_t \choose k_t} {n_r \choose k_r}\right)$ is a constant independent of SNR and channel coefficients.
    \label{constant_gap}
    
    Moreover, there exist MIMO channel configurations for which $C_{k_t,k_r}^{\max} = \frac{\min(k_t,k_r)}{\min(n_t,n_r)}\ C$.
\end{theorem}

%% file: proof.tex
\section{Proofs}
\emph{Note:} Throughout the remainder of the paper, we use $[n]$ to denote a set of integers from 1 to $n$.

To prove the theorems in Section \ref{mainresults}, our arguments borrow tools from Linear Algebra, particularly, the following property on principal submatrices.
\begin{property}\label{property_submatrices}
    Let $\mb{A}$ be an $n \times n$ Hermitian matrix and let $\Lambda \subseteq [n]$, where $|\Lambda|=k$.
    Define $\mb{A}_\Lambda$ to be the submatrix of $\mb{A}$, constructed only from the rows and columns of $\mb{A}$ indexed by $\Lambda$.
    Let $\rho(\lambda)$ and $\rho_\Lambda(\lambda)$ be the characteristic polynomials of $\mb{A}$ and $\mb{A}_\Lambda$, respectively.
    Then the following property holds:
    \begin{equation}
        \label{rho_2_rhosub_gn}
        (n-k)!\sum_{\substack{\Lambda \subseteq [n]\\|\Lambda|=k}}\rho_{\Lambda}(\lambda)= \rho^{(n-k)}(\lambda),
    \end{equation}
    where: (i) the summation in \eqref{rho_2_rhosub_gn} is over all subsets of $[n]$ of cardinality $k$; (ii) $f^{(j)}(x)$ is the $j$-th derivative of $f(x)$ with respect to $x$.
\end{property}

Property \ref{property_submatrices} is mentioned in \cite{thompson} as a well-known fact.
For completeness, we include a simple proof of the property in Appendix~\ref{appendixA} based on the multilinearity of determinants.

\emph{Remark}: Applying \eqref{rho_2_rhosub_gn} with $k=n-1$ leads to the following identity:
\begin{equation}
    \sum_{i=1}^n \rho_{[n]\setminus i}(\lambda) = \rho^{(1)}(\lambda),
    \label{property_n-1}
\end{equation}
which we use in the proof of Theorem \ref{zero_gap_thm}.
\subsection{Proof of Theorem~\ref{zero_gap_thm}}
To prove the lower bound in Theorem \ref{zero_gap_thm}, it suffices to prove the statement for the following two incremental cases:
\begin{enumerate}
    \item For $k_t = n_t,k_r = n_r-1$, $C_{n_t,n_r-1}^\star \geq \frac{n_r-1}{n_r} C$,\\
    \item $k_t = n_t-1,k_r = n_r$, $C_{n_t-1,n_r}^\star \geq \frac{n_t-1}{n_t} C$.
\end{enumerate}
The above two statements would imply that we can reduce an $n_t \times n_r$ system to a $k_t \times k_r$ system as follows: We first remove one receiver antenna to create an $n_t \times (n_r - 1)$ system such that its capacity $C_{n_t, n_r - 1}^\star \geq \frac{n_r - 1}{n_r}C_{n_t,n_r} = C$. From this (particular) $n_t \times (n_r - 1)$ system, we select an $n_t \times (n_r - 2)$ system such that its capacity $C_{n_t, n_r - 2}^\star \geq \frac{n_r - 2}{n_r -1}C_{n_t,n_r-1}^\star$, and so on, till we prune the system down to a $n_t \times k_r$ system. We then repeat the above process for transmitter selection on the $n_t \times k_r$ system to prune it progressively to a $k_t \times k_r$ system with capacity $C_{k_t,k_r}^\star$ , The result would then follow as:
\[
\begin{aligned}
C^\star_{k_t,k_r} \geq& \quad \dfrac{k_t}{k_t + 1} C^\star_{k_t+1,k_r}\\
			 \geq& \quad \dfrac{k_t}{k_t + 1} \dfrac{k_t+1}{k_t + 2}C^\star_{k_t+2,k_r}\\
			 \geq& \quad \dfrac{k_t}{k_t + 1} \dfrac{k_t+1}{k_t + 2} .. \dfrac{n_t-1}{n_t}\ C^\star_{n_t,k_r}\\
             \geq& \quad \dfrac{k_t}{k_t + 1} \dfrac{k_t+1}{k_t + 2} .. \dfrac{n_t-1}{n_t}\dfrac{k_r}{k_r+1}\ C^\star_{n_t,k_r+1}\\
             \geq& \quad \dfrac{k_t}{n_t}\dfrac{k_r}{k_r+1} .. \dfrac{n_r-1}{n_r}\ C_{n_t,n_r}\\			 			 			
             \geq& \quad \dfrac{k_t k_r}{n_t n_r}\ C_{n_t,n_r}\\			
\end{aligned}
\]
where $C_{n_t,n_r} = C$ is the capacity of the full $n_t \times n_r$ MIMO channel.
\\
Since we can rewrite $C = \log\det(\mathbf{I} + P\mathbf{H}\mathbf{H}^\dagger)$ as $\log\det(\mathbf{I} + \tilde{\mathbf{H}}\tilde{\mathbf{H}}^\dagger)$ where $\tilde{\mathbf{H}} = \sqrt{P}\mathbf{H}$, without loss of generality, we will subsequently assume that the power constraint $P$ is unity, as proving the Theorem for $P \neq 1$ is equivalent to proving it for $\tilde{\mathbf{H}}$ instead of $\mathbf{H}$.
\\

\noindent \textit{\textbf{Case 1: ($k_t = n_t$, $k_r = n_r-1$)}}\\
Let $\mathbf{F} = \mathbf{I}+\mathbf{H}\mathbf{H}^\dagger$. The capacity can then be written as $C = \log\det(\mathbf{F})$.
We define $ \mathbf{H}_{[n_r]\setminus i}$ to be the submatrix of $\mathbf{H}$ constructed by dropping the $i$-th receiver antenna ($i$-th row in $\mathbf{H}$).
Let $\mathbf{B}_i = \mathbf{I} + \mathbf{H}_{[n_r]\setminus i} \mathbf{H}^\dagger_{[n_r] \setminus i}$. Therefore, $C_i = \log\det(\mathbf{I} + \mathbf{H}_{[n_r]\setminus i} \mathbf{H}^\dagger_{[n_r] \setminus i}) = \log\det(\mathbf{B}_i)$ is the capacity of the MIMO system with the remaining $n_r- 1$ receiver antennas.
\\

\noindent Let $\rho(\lambda)$ denote the characteristic polynomial of $\mathbf{F}$ and let $\rho_{[n_r]\setminus i}(\lambda)$ denote the characteristic polynomial for $\mathbf{B}_i$ respectively. These polynomials can be represented as:
\[
    \rho(\lambda) = \sum_{j=0}^{n_r} f_{(j)}\lambda^{j}
\]
\[
    \rho_{[n_r]\setminus i}(\lambda) = \sum_{j=0}^{n_r-1} b_{(i,j)}\lambda^{j}
\]
where $f_{(n_r)}= 1$ and $b_{(i,n_r-1)} = 1 $ for all $i \in [n_r]$.

Plugging this into \eqref{property_n-1} yields:
    \begin{align}
        \label{expanded_eqn} \sum_{j=0}^{n_r-1}\sum_{i=1}^{n_r} b_{(i,j)}\lambda^{j} =& \sum_{j=0}^{n_r-1} (j+1)f_{(j+1)}\lambda^{j}
    \end{align}
Comparing the coefficients of $\lambda^0$ in \eqref{expanded_eqn} we get:
\begin{equation}\label{diff_rho}
     \sum_{i=1}^{n_r} b_{(i,0)} = f_{(1)}
\end{equation}

Note that for any positive semidefinite matrix, the characteristic polynomial of order $n$ can be factorized into the form:
\begin{align}
 \rho(\lambda) =   (\lambda - \hat{\lambda}_1)(\lambda - \hat{\lambda}_2) .. (\lambda - \hat{\lambda}_n)
    \label{eq:lambda_factorization}
\end{align}
where $\{\hat{\lambda}_1,\hat{\lambda}_2,\dots,\hat{\lambda}_{n_r}\}$ are the eigenvalues of the matrix.
Using the factorization above, $f_{(1)}$ can be written as:
    \begin{equation}
        f_{(1)} = (-1)^{n_r-1} \sum_{\substack{\Gamma \subset [n_r]\\|\Gamma|=n_r-1}}\left(\ \prod_{i \in \Gamma} \lambda_{i} \right)
        \label{f_1_expanded}
    \end{equation}
where $\{\lambda_1,\lambda_2,\dots,\lambda_{n_r}\}$ are the eigen values of the matrix $\mathbf{F}$ and the summation in \eqref{f_1_expanded} is over all $(n_r -1)$-tuples of the eigenvalues of $\mathbf{F}$.

Therefore, from \eqref{diff_rho} and \eqref{f_1_expanded} (and dividing both sides by $n_r$), we have:
\begin{equation}
\begin{aligned}
    \dfrac{(-1)^{n_r-1}}{n_r} \sum_{i=1}^{n_r} b_{(i,0)} =&  \dfrac{1}{n_r} \sum_{\substack{\Gamma \subseteq [n_r]\\|\Gamma|=n_r -1}}\left(\ \prod_{i \in \Gamma} \lambda_{i} \right)\\
    \stackrel{(a)}{\geq}&  \left\{\prod_{\substack{\Gamma \subseteq [n_r]\\|\Gamma|=n_r -1}}\left(\ \prod_{i \in \Gamma} \lambda_{i} \right)\right\}^\frac{1}{n_r}\\
    \stackrel{(b)}{=}& \prod_{i=1}^{n_r} \lambda_i^\frac{n_r - 1}{n_r}
\end{aligned}
    \label{AMGM_counting}
\end{equation}
where (a) follows from the AM-GM inequality and (b) follows since in all $(n_r -1)$-tuples of eigenvalues, any particular eigenvalue appears as part of exactly ${n_r-1 \choose n_r-2} = (n_r -1)$-tuples.

Using the factorization of the characteristic polynomial in \eqref{eq:lambda_factorization}, we can also express the term $\rho_{[n_r]\setminus i}(\lambda = 0) = (-1)^{n_r-1} b_{(i,0)}$ as the product of all eigenvalues of $\mathbf{B}_i$ (and hence its determinant). As a result, we can write \eqref{AMGM_counting} as:
\begin{equation}
    \dfrac{1}{n_r} \sum_{i=1}^{n_r} \det\left( \mathbf{B}_i \right)   \geq \prod_{i=1}^{n_r} \lambda_i^\frac{n_r - 1}{n_r} = \det\left(\mathbf{F} \right)^\frac{n_r-1}{n_r}
    \label{det_B_det_F}
\end{equation}
Since the left hand side is average of the determinants of $\mathbf{B}_i$, $i \in [n_r]$, \eqref{det_B_det_F} implies one of following:
\begin{enumerate}
    \item $\exists\ i_+,i_-  \in [n_r]$ such that
    \[
        \det\left(\mathbf{B}_{i_+}\right) > \det\left(\mathbf{F} \right)^\frac{n_r-1}{n_r},\quad \det\left(\mathbf{B}_{i_-} \right) < \det\left(\mathbf{F} \right)^\frac{n_r-1}{n_r}
    \]
\item $\det\left(\mathbf{B}_i \right) \geq \det\left(\mathbf{F}\right)^\frac{n_r-1}{n_r}  \quad \forall i \in \{1,2,...,n\}$
\end{enumerate}
Both cases imply that there exists some selection of  $n_r - 1$ receivers (by removing the receiver $i_\star$) such that:
\[
    C^\star_{n_t,n_r-1} = \log \det\left(\mathbf{B}_{i_*} \right) \geq \log \det\left(\mathbf{F}\right)^\frac{n_r-1}{n_r}
\]
Since $C = \log\det\left( \mathbf{F} \right)$, we have:
\[
    C^\star_{n_t,n_r-1} \geq \dfrac{n_r-1}{n_r}\ C
\]
This concludes the proof for the first case.
\\

\noindent \textit{\textbf{Case 2: ($k_t = n_t-1$, $k_r = n_r$)}}\\
To prove this case, we appeal to Sylvester's determinant theorem that states that
\[
    C = \log\det(\mathbf{I}_{n_r} + \mathbf{H} \mathbf{H}^{\dagger}) = \log\det(\mathbf{I}_{n_t} + \mathbf{H}^{\dagger}\mathbf{H}).
\]
Let $\hat{\mathbf{F}} = \mathbf{I}_{n_t} + \mathbf{H}^\dagger \mathbf{H}$, and therefore, $C = \log\det(\hat{\mathbf{F}})$.
We denote by ${\mathbf{H}^\dagger}_{[n_t]\setminus j}$, the submatrix of $\mathbf{H}^\dagger$ after dropping the $j$-th row.
The capacity of this MIMO subchannel can also be written by Sylvester's theorem as $C_j = \log\det\left(\mathbf{I}_{n_t} + {\mathbf{H}^\dagger}_{[n_t]\setminus j} ({\mathbf{H}^\dagger}_{[n_t]\setminus j})^\dagger  \right) = \log\det(\hat{\mathbf{B}}_j)$ where $\hat{\mathbf{B}}_j$ is the $(n_t -1) \times (n_t -1)$ matrix constructed from $\hat{\mathbf{F}}$ after removing the $j$-th column and row.
The argument to prove the ratio $\frac{n_t-1}{n_t}$ thus follows similarly as in Case 1 with $\mathbf{B}_i$ and $\mathbf{F}$.
\\
\paragraph*{Tight Example}
To prove that the lower bound in Theorem \ref{zero_gap_thm} is tight, consider the $n_t \times n_r$ MIMO channel described by $\mb{H} =\sqrt{P} \mb{O}_{n_r,n_t}$, where $\mb{O}_{n_r,n_t}$ is a $n_r \times n_t$ matrix with all entries equal to unity. It is not hard to see that for the described channel,
\[
    C = \log(1+P n_t n_r).
\]
Similarly for any subchannel of size $k_t \times k_r$, the capacity is $C_{k_t,k_r} = \log(1+P k_t k_r)$.
Note that for $x \approx 0$, we have $\log(1+x) \approx \frac{1}{\ln(2)}x$. Therefore for $P \approx 0$, we get that $C \approx \frac{1}{\ln(2)} P n_t n_r$ and similarly $C_{k_t,k_r} \approx  \frac{1}{\ln(2)} P k_t k_r$. 
Therefore for $P \approx 0$,
\[
    \frac{C_{k_t,k_r}}{C} \approx \frac{k_t k_r}{n_t n_r}.
\]
\noindent This concludes our proof of Theorem \ref{zero_gap_thm}.

\subsection{Proof of Theorem \ref{const_gap_thm}}
Let $\mathbf{F} = \mathbf{I} + \mathbf{H}\mathbf{H}^\dagger$ and define $\lambda_1 \geq \lambda_2 \geq \dots \geq \lambda_{n_r}$ to be the eigenvalues of $\mathbf{F}$.
To prove theorem \ref{const_gap_thm}, we appeal to the the property of characteristic polynomials described in \eqref{rho_2_rhosub_gn}.
For our purposes, the variables $n$ and $k$ in \eqref{rho_2_rhosub_gn} are replaced with $n_r$ and $k_r$, respectively to give the following:
\begin{align}
    (n_r-k_r)!\sum_{\substack{\Lambda_i \in \Pi }}\rho_{\Lambda_i}(\lambda)= \rho^{(n_r-k_r)}(\lambda),
    \label{eq:relation_edited}
\end{align}
where $\Pi$ is the set of all unique subsets $\Lambda_i \subseteq [n_r]$, $|\Lambda_i| = k_r$.
By comparing the coefficents of $\lambda^0$ in \eqref{eq:relation_edited}, we have:
\[
    (n_r - k_r)!\sum_{i=1}^{|\Pi_{r}|} b_{(i,0)}   = (n_r - k_r)! f_{(n_r - k_r)}
\]
Using \eqref{eq:lambda_factorization}, we can write the coefficient $f_{(n_r - k_r)}$ as:
\[
    f_{(n_r-k_r)}\ \ = \sum_{\substack{\{j_1,...,j_{k_r}\} \subseteq [n_r]}}\hspace{-0.1in}\lambda_{j_1}\lambda_{j_2}...\lambda_{j_{k_r}}
\]
which is the sum of the (product of) eigenvalues of $\mathbf{F}$, taken $k_r$ at a time.
Therefore we have :
\begin{equation}
    \sum_{i=1}^{|\Pi_r|} b_{(i,0)}   = \sum_{\substack{\{j_1,...,j_{k_r}\} \subseteq [n_r]}}\hspace{-0.1in}\lambda_{j_1}\lambda_{j_2}...\lambda_{j_{k_r}}
     \label{thm2_main_eqn}
 \end{equation}
 Without loss of generality, and to simplify notation, we shall assume throughout this subsection that $n_t = \min(n_t,n_r)$.
 Since we have $n_t \leq n_r$, there exists at most $n_t$ eigenvalues of $\mathbf{F} = \mathbf{I} + \mathbf{H}\mathbf{H}^\dagger$ that are not equal to unity, i.e.,  $\lambda_i = 1$ for $i \in \{ n_t+1,n_t+2,\cdots n_r\}$
 We shall prove the theorem for two incremental cases and then show recursively, that the theorem holds for all other cases.
 The base cases we need to prove are the following:
\begin{enumerate}
    \item For $k_r \leq n_t \leq n_r$, there exists a MIMO subchannel of dimensions $n_t \times k_r$ and capacity $C^\star_{n_t,k_r}$ such that
        \begin{equation}
            C^\star_{n_t,k_r} \geq \frac{k_r}{n_t}\ C - \log\left(\frac{{n_r \choose k_r}}{{n_t \choose k_r}}\right).
            \label{k_r_smaller_n_t}
        \end{equation}
    \item For $n_t \leq k_r \leq n_r$, there exists a MIMO subchannel of dimensions $n_t \times k_r$ and capacity $C^\star_{n_t,k_r}$ such that
        \begin{equation}
            C \geq\  C^\star_{n_t,k_r} \geq\ C - \log\left(\frac{{n_r \choose k_r}}{{n_r-n_t \choose k_r-n_t}}\right).
            \label{k_r_larger_n_t}
        \end{equation}
\end{enumerate}
We can combine the lower bounds in \eqref{k_r_smaller_n_t} and \eqref{k_r_larger_n_t} as
\begin{align}
    C^\star_{k_t,k_r} \geq \frac{\min(k_r,n_t)}{n_t}\ C - G,
    \label{eq:combined_cases_const_thm}
\end{align}
where $G$ is the constant incurred in \eqref{k_r_smaller_n_t} (resp. \eqref{k_r_larger_n_t}) when $k_r \leq n_t$ (resp. $k_r > n_t$).
Note that \eqref{eq:combined_cases_const_thm} applies similarly for the case when $n_r = \min(n_t,n_r)$ simply by considering the reciprocal MIMO channel or appealing to Sylvester's determinant theorem.
Using \eqref{eq:combined_cases_const_thm}, we can now derive the bound on $C^\star_{k_t,k_r}$ for any chosen dimension $(k_t,k_r)$ as follows: 
From the $n_t \times n_r$ channel, we can create (applying \eqref{eq:combined_cases_const_thm}) an $n_t \times k_r$ subchannel such that $C^\star_{n_t,k_r} \geq \frac{\min(k_r,n_t)}{n_t}\ C_{n_t,n_r} - G_1$, by keeping only the best $k_r$ receiver antennas. Next from this $n_t \times k_r$ channel, we can again get a $n_t \times n_r$ subchannel such that
\begin{align*}
    C^\star_{k_t,k_r} &\geq \frac{\min(k_t,k_r)}{\min(n_t,k_r)}\ C^\star_{n_t,k_r} - G_2 \\
    &\geq \frac{\min(k_t,k_r)}{\min(n_t,n_r)}\ C_{n_t,n_r} - G_1 - G_2.
\end{align*}
In particular, the constants $G_1$ and $G_2$ are captured in the following three cases:
\begin{enumerate}
    \item For $k_t \leq k_r \leq n_t \leq n_r$:
        \begin{align*}
                C^\star_{k_t,k_r} \stackrel{(a)}\geq& \frac{k_t}{k_r}\ C^\star_{n_t,k_r} - \log\left(\frac{ {n_t \choose k_t } }{ {k_r \choose k_t } } \right)\\
                \stackrel{(b)}\geq& \frac{k_r}{n_t}\frac{k_t}{k_r} C_{n_t,n_r} - \frac{k_t}{k_r}\log\left(\frac{ {n_r \choose k_r } }{ {n_t \choose k_r } } \right) - \log\left(\frac{ {n_t \choose k_t } }{ {k_r \choose k_t } } \right)\\
                \geq& \frac{k_t}{n_t}\ C - \log\left( {n_t \choose k_t }  \right) -\log\left( {n_r \choose k_r } \right),
        \end{align*}
        where: (a) follows by applying \eqref{k_r_smaller_n_t} on the reciprocal of the MIMO channel $n_t \times k_r$; (b) applies \eqref{k_r_smaller_n_t} to relate  $C^\star_{n_t,k_r}$ to $C_{n_t,n_r}$.\\
    \item For $k_r \leq k_t \leq n_t \leq n_r$:
        \[
            \begin{aligned}
                C^\star_{k_t,k_r} \stackrel{(c)}\geq&\ C^\star_{n_t,k_r} - \log\left(\frac{ {n_t \choose k_t } }{ {n_t-k_r \choose k_t-k_r } } \right)\\
                \stackrel{(d)}\geq& \frac{k_r}{n_t}\ C_{n_t,n_r} - \log\left(\frac{ {n_t \choose k_t } }{ {n_t-k_r \choose k_t-k_r } } \right) -\log\left(\frac{ {n_r \choose k_r } }{ {n_t \choose k_r } } \right)\\
                \geq& \frac{k_r}{n_t}\ C - \log\left( {n_t \choose k_t }  \right) -\log\left( {n_r \choose k_r } \right),
            \end{aligned}
        \]
        where: (c) relates $C_{k_t,k_r}$ to $C_{n_t,k_r}$ using \eqref{k_r_larger_n_t}; (d) follows by applying \eqref{k_r_smaller_n_t} on the $n_t \times n_r$ MIMO channel.\\
    \item For $k_t \leq n_t \leq k_r \leq n_r$:
        \[
            \begin{aligned}
                C^\star_{k_t,k_r} \stackrel{(e)}{\geq}&\ \frac{k_t}{n_t}\ C^\star_{n_t,k_r}\\
                \stackrel{(f)}\geq& \frac{k_t}{n_t}\ C_{n_t,n_r} - \log\left(\frac{ {n_r \choose k_r } }{ {n_r-n_t \choose k_r-n_t } } \right)\\
                \geq& \frac{k_t}{n_t}\ C - \log\left( {n_t \choose k_t }  \right) -\log\left( {n_r \choose k_r } \right),
            \end{aligned}
        \]
        where (e) follows by applying Theorem \ref{zero_gap_thm} to select an $k_t \times k_r$ subchannel from the $n_t \times k_r$ MIMO channel; The relation (f) follows from \eqref{k_r_larger_n_t}.
\end{enumerate}
By combining the aforementioned cases, we have:
\[
    C^\star_{k_t,k_r} \geq \frac{\min{(k_t,k_r)}}{\min{(n_t,n_r)}}\ C - \log\left( {n_t \choose k_t }  {n_r \choose k_r } \right)
\]

Now to conclude the proof, we need to assert the bounds for the two cases in \eqref{k_r_smaller_n_t} and \eqref{k_r_larger_n_t} respectively.\\

\noindent \textit{\textbf{Case 1: ($k_r \leq n_t \leq n_r$)}}\\
The expression in \eqref{thm2_main_eqn} can be simplified when $k_r \leq n_t$ as follows:
\begin{equation} \label{const_gap_case_1}
\begin{aligned}
    \sum_{i=1}^{|\Pi_{k_r}|} b_{(i,0)}   =&  \sum_{\{j_1,...,j_{k_r}\}  \subseteq [n_r]}\hspace{-0.1in}\lambda_{j_1}\lambda_{j_2}...\lambda_{j_{k_r}} \\
    \stackrel{(a)}{\geq}& {n_t \choose k_r} \sum_{\{j_1,...,j_{k_r}\} \subseteq [n_t]}\hspace{-0.1in}\dfrac{\lambda_{j_1}\lambda_{j_2}...\lambda_{j_{k_r}}}{{n_t \choose k_r}} \\
    \stackrel{(b)}{\geq}& {n_t \choose k_r}\mathlarger{\prod}_{\{j_1,...,j_{k_r}\} \subseteq [n_t]}\left(\lambda_{j_1}\lambda_{j_2}...\lambda_{j_{k_r}}\right)^{{n_t \choose k_r}^{-1}}\\
    =& {n_t \choose k_r}\left(\mathlarger{\prod}_{i=1}^{n_t}\lambda_{i}\right)^{{n_t-1 \choose k_r-1}{n_t \choose k_r}^{-1}}\\
    \stackrel{(c)}{=}& {n_t \choose k_r}\left(\mathlarger{\prod}_{i=1}^{n_r}\lambda_{i}\right)^{\frac{k_r}{n_t}}\\
\end{aligned}
\end{equation}

where (a) by considering only $k_r$-tuples of the eigen values $\lambda_i$ where $i \in [n_t]$.
Since $[n_t] \subseteq [n_r]$, then all $k_r$-tuples from $[n_t]$ are contained within the summation in \eqref{thm2_main_eqn} and therefore the relation follows.
The relation (b) follows from the AM-GM inequality.
(c) follows by the simplification of the exponent and the fact that $\lambda_i = 1$ for $i \in \{n_t+1,\dots n_r\}$.

By averaging the left hand side of \eqref{const_gap_case_1}, we have:
\[
    \dfrac{1}{{n_r \choose k_r}}\sum_{i=1}^{\vert \Pi_r \vert} \det(\mathbf{B}_i) = \dfrac{1}{{n_r \choose k_r}}\sum_{i=1}^{\vert \Pi_r \vert} b_{(i,0)} \geq \frac{{n_t \choose k_r}}{{n_r \choose k_r}} \left(\mathlarger{\prod}_{i=1}^{n_r}\lambda_{i}\right)^{\frac{k_r}{n_t}}
\]
This implies that there exists some selection $\Lambda_*$ of $k_r$ receivers such that $\mathbf{B}_* = \mathbf{I} + \mathbf{H}_{\Lambda_*}{\mathbf{H}_{\Lambda_*}}^\dagger$ and we have:
\[
    \begin{aligned}
        \log\det(\mathbf{B}_*) \geq& \log \left(\frac{{n_t \choose k_r}}{{n_r \choose k_r}} \left(\mathlarger{\prod}_{i=1}^{n_r}\lambda_{i}\right)^{\frac{k_r}{n_t}} \right) \\
        =& \log\det(\mathbf{F})^{\frac{k_r}{n_t}} - \log\left(\frac{{n_r \choose k_r}}{{n_t \choose k_r}}\right)
    \end{aligned}
\]

As a result, the capacity of the best MIMO subchannel from choosing $k_r$ receivers out of $n_r$, where $k_r \leq n_t$ is:
\[
    C^\star_{n_t,k_r} \geq \frac{k_r}{n_t} C - \log\left( \frac{{n_r \choose k_r}}{{n_t \choose k_r}}  \right)
\]
\noindent \textit{\textbf{Case 2: ($k_t = n_t$, $n_t \leq k_r \leq n_r$)}}\\

Since $k_r \geq n_t$, there exist $k_r$-tuples in \eqref{thm2_main_eqn} such that $[n_t] \subseteq \{j_1,\dots,j_{k_r}\} \subseteq [n_r]$.
There are  ${n_r-n_t \choose k_r-n_t}$ such tuples and therefore, we have:

\begin{equation} \label{const_gap_case_2_LB}
\begin{aligned}
    \dfrac{1}{{n_r \choose k_r}}\sum_{i=1}^{\vert \Pi_r \vert} \det(\mathbf{B}_i) =& \dfrac{1}{{n_r \choose k_r}}\sum_{i=1}^{\vert \Pi_r \vert} b_{(i,0)}   \\
    =& \dfrac{1}{{n_r \choose k_r}} \sum_{\{j_1,...,j_{k_r}\}  \subseteq [n_r]}\hspace{-0.2in}\lambda_{j_1}\lambda_{j_2}...\lambda_{j_{k_r}} \\
    \stackrel{(a)}{\geq} & \dfrac{{n_r-k_r \choose k_r-n_t}}{{n_r \choose k_r}} \left(\mathlarger{\prod}_{i=1}^m \lambda_i \right)
\end{aligned}
\end{equation}

The relation in  \eqref{const_gap_case_2_LB} implies that there exits a selection $\Lambda_*$ of $k_r$-receivers such that:
\[
    \log\det(\mathbf{B}_*)  \geq \log\det(\mathbf{F}) - \log\left(\frac{{n_r \choose k_r}}{{n_r-k_r \choose k_r-n_t}}\right)
\]
Therefore the best subchannel by choosing $k_r$ receivers is:
\[
    C^\star_{n_t,k_r} \geq C - \log\left( \frac{{n_r \choose k_r}}{{n_r-n_t \choose k_r-n_t}}  \right)
\]
However, fundamentally, $C^\star_{n_t,k_r} \leq C$, therefore, we have:
\[
    C \geq C^\star_{n_t,k_r} \geq  C - \log\left( \frac{{n_r \choose k_r}}{{n_r-n_t \choose k_r-n_t}}  \right)
\]

\noindent This concludes the proof of the lower bound in Theorem \ref{const_gap_thm}.

\paragraph*{Tight Example}
To prove that there exists a class of networks for which the lower bound in Theorem \ref{const_gap_thm} is tight (to within a constant gap), consider the $n \times n$ MIMO channel described by $\mb{H}= \sqrt{P}\mb{I}$.
This is a parallel MIMO channel, where each of the individual parallel channels is of capacity $\log(1+P)$ and the capacity of the full network is $C = n\log(1+P)$.
For any $(k_t,k_r)$, it is not hard to see that a $k_t \times k_r$ MIMO subchannel can at most capture $\min(k_t,k_r)$ of the parallel channels. Therefore, we have $C^\star_{k_t,k_r} = \min(k_t,k_r)\log(1+P)$ and as a result
\[
    \frac{C^\star_{k_t,k_r}}{C} = \frac{\min(k_t,k_r)}{n}.
\]

%% file: appendices.tex
\begin{appendices}
    \section{Proof of Property \ref{property_submatrices}} \label{appendixA}
    \noindent Let $\rho(\lambda)$ denote the characteristic polynomial of matrix $\mathbf{A}$.
    The characteristic polynomial $\rho(\lambda)$ is equal to the determinant of $(\lambda \mathbf{I} - \mathbf{A})$ and is therefore, by the property of determinants, multilinear in the rows of the matrix $\lambda \mathbf{I} - \mathbf{A}$.
    This means we can write $\rho(\Lambda)$ as :
    \[
        \rho(\lambda) = M\big(r_1(\lambda),r_2(\lambda), ... r_n(\lambda)\big)
    \]
    where $M : \mathbb{C}^n \times \mathbb{C}^n \cdots \times \mathbb{C}^n \rightarrow \mathbb{R}$ is a multilinear mapping and $r_i(\lambda)$ is the $i$-th row of $\mathbf{A}$.
    Since M is multilinear, its total derivative is the sum of its partial derivatives \cite{spivak} ,i.e.,
    \[
        M^{(1)}(x_1,x_2,..,x_n).(y_1,y_2,..,y_n) = \sum_{i=1}^n M(x_1,...,y_i,..,x_n)
    \]
    Therefore, by applying the chain rule, we have:
    \[
        \begin{aligned}
            \rho^{(1)}(\lambda) &=  M(x_1,x_2,..,x_n).(r_1^{(1)}(\lambda),r_2^{(1)}(\lambda),..,r_n^{(1)}(\lambda))\\
            &= \sum_{i=1}^n M\big(r_1(\lambda), ... , r^{(1)}_i(\lambda), ... r_n(\lambda)\big)
        \end{aligned}
    \]
    where $r^{(1)}_i(\lambda)$ is the differentiation of the $i$-th row of $\mathbf{A}$ with respect to $\lambda$.
    Therefore $r^{(1)}(\lambda) = 0$ at all non-diagonal positions and equals 1 at the diagonal position.
    $M\big(r_1(\lambda), ... , r^{(1)}_i(\lambda), ... r_n(\lambda)\big)$ is the determinant of the matrix $\lambda \mathbf{I} - \mathbf{A}$ after replacing the $i$-th row by $r'_i(\lambda)$.
    Expanding the determinant along the $i$-th row of this new matrix, we get that:
    \[
        M\big(r_1(\lambda), ... , r^{(1)}_i(\lambda), ... r_n(\lambda)\big) = 1\times(\lambda \mathbf{I}-\mathbf{A})_{ii}
    \]
    where $(\lambda \mathbf{I}-\mathbf{A})_{ii}$ is the minor of $\lambda \mathbf{I}-\mathbf{A}$ formed by removing the $i$-th row and $i$-th column, which is equal to $\det(\lambda \mathbf{I} + \mathbf{A}_{[n]\setminus i})$.
    $\mathbf{A}_{[n]\setminus i}$ is the submatrix of $\mathbf{A}$ by removing the $i$-th row and $i$-th column.
    As a result, we have:
    \begin{equation}
        \rho^{(1)}(\lambda) = \sum_{i=1}^n \det(\lambda \mathbf{I} + \mathbf{A}_{[n]\setminus i}) = \sum_{i=1}^n \rho_{[n]\setminus i}(\lambda).
        \label{rho_2_rhosub}
    \end{equation}
    where $\rho_{[n]\setminus i}(\lambda)$ denotes the characteristic polynomial of $\mathbf{A}_{[n]\setminus i}$ and $i \in \{1,2,\cdots n\}$.
    \\
    To prove the relation in \eqref{rho_2_rhosub_gn}, we need an induction relation in addition to \eqref{rho_2_rhosub}

    Let $g_{k+1}(\lambda)$ be the sum of all characteristic equations of $k+1 \times k+1$ submatrices, i.e.,
    \[
        g_{k+1}(\lambda) = \sum_{\substack{\Lambda \subseteq [n]\\|\Lambda|=k+1}} \rho_{\Lambda}(\lambda)
    \]

    Taking the derivative of $g_{k+1}(\lambda)$ and applying \eqref{rho_2_rhosub}, we get:
    \begin{equation}
        g^{(1)}_{k+1}(\lambda) =  \sum_ {\substack{\Lambda \subseteq [n]\\|\Lambda|=k+1}} \rho^{(1)}_{\Lambda}(\lambda)
        = \sum_{\substack{\Lambda \subseteq [n]\\|\Lambda|=k+1}} \sum_{j \in \Lambda} \rho_{\Lambda\setminus j}(\lambda)
        \label{diff_k_plus_1}
    \end{equation}
    where $\rho_{\Lambda\setminus j}(\lambda)$ is the characteristic polynomial of the $k\times k$ submatrix of $\mathbf{A}$ with rows and columns in $\Lambda\setminus j$.
    Since there are only ${n \choose k}$ submatrices of size $k \times k$, the summation in \eqref{diff_k_plus_1} is bound to have repeated terms.
    By a simple counting argument, we can see that for each matrix, there are $n-k$ copies of its characteristic polynomial in \eqref{diff_k_plus_1}.
    This can be observed by noting that in \eqref{diff_k_plus_1}, the inner summation consists of $k+1$ terms and the other summation is over ${n \choose k}$ terms.
    It is easy to verify that:
    \[
        {n \choose k+1}(k+1) = {n \choose k}(n-k)
    \]
    As a result, we can write \eqref{diff_k_plus_1} as:
    \begin{equation}
        \sum_ {\substack{\Lambda \subseteq [n]\\|\Lambda|=k+1}} \rho^{(1)}_{\Lambda}(\lambda) = (n-k) \sum_ {\substack{\Lambda \subseteq [n]\\|\Lambda|=k}} \rho_{\Lambda}(\lambda)
        \label{induction_step}
    \end{equation}
    which is our induction hypothesis.

    Our base case is what we proved in \eqref{rho_2_rhosub} which can be deduced from \eqref{induction_step} by choosing $k = n-1$:
    Therefore, by induction, we get:
    \begin{equation}
        (n-k)!      \sum_ {\substack{\Lambda \subseteq [n]\\|\Lambda|=k}} \rho_{\Lambda}(\lambda) = \rho^{(n-k)}(\lambda)
    \end{equation}

\end{appendices}